\documentclass{article}
\usepackage{varioref}
\usepackage{amssymb}
\usepackage{color}
\usepackage{amsmath}
\usepackage{graphicx}

\begin{document}

\title{\bf Dynamical Casimir Effect and Vacuum Friction in the Near-Horizon Geometry of a Black Hole}

\author{
\begin{tabular}{c}
H. Hadi\thanks{email: hamedhadi1388@gmail.com} \\
{\small Faculty of Physics, University of Tabriz, Tabriz 51666-16471, Iran} \\[1.2ex]
Amin Rezaei Akbarieh\thanks{email: amin.rezaeiakbarieh@kocaeli.edu.tr} \\
{\small Department of Physics, Kocaeli University, 41001 Izmit, T\"urkiye}\\[1.2ex]
G\"oksel Daylan Esmer\thanks{Email: goksel@istanbul.edu.tr}\\
{\small Department of Physics, Istanbul University, Vezneciler, 34134 Istanbul, T\"urkiye}
\end{tabular}
}

\date{\today}

\maketitle

\begin{abstract}
We investigate the Dynamical Casimir Effect (DCE) for a relativistic scalar field confined within a cavity possessing moving boundaries in the (1+1)-dimensional near-horizon geometry of a black hole. By applying a coordinate transformation, we map the moving-boundary problem to an equivalent acoustic metric with static boundaries, allowing for an exact canonical Hamiltonian formulation. We find that the local gravitational redshift fundamentally alters the vacuum structure, and the dynamical boundary motion induces  time-dependent mode-mixing. When a boundary moves, it scatters the fluctuations of the ambient Hartle-Hawking state, generating a flux of created particles. Crucially, because the coordinate speed of light relative to the Killing time $t$ vanishes  as one approaches the event horizon, we establish that maintaining physical, subluminal boundary motion requires the mechanical oscillation amplitude to scale proportionally with the proper distance to the horizon. Consequently, the effective Mach number of the moving mirror  approaches zero in the near-horizon limit. Using a rigorous small-amplitude perturbative expansion and proper canonical operator normalization, we demonstrate that the transition probability into the field is heavily suppressed by a conformal geometric factor. Furthermore, we account for the Bose-enhancement caused by the thermal Hawking bath. While the thermal presence introduces infrared density-of-states enhancement, it remains insufficient to overcome the kinematic damping. Finally, we conclude that the extreme spacetime curvature acts to protect the near-horizon vacuum; the transition probability vanishes as the boundary approaches the event horizon, indicating a  geometric and kinematic suppression of particle creation in the strong-gravity limit.
\end{abstract}

\section{Introduction}

The confinement of a quantum field within a cavity subject to perfectly reflecting boundaries has long served as a fundamental laboratory for investigating the non-trivial structure of the quantum vacuum. When the boundaries are held fixed, the restriction of vacuum fluctuations yields the well-established static Casimir effect \cite{Milton2001}. Extending this setup to configurations where one or both boundaries execute time-dependent motions defines the Dynamical Casimir Effect (DCE) \cite{10}, a phenomenon wherein the mechanical motion of the boundaries non-adiabatically scatters vacuum fluctuations, converting them into real, on-shell particles. 

The most thoroughly examined prototypes of the DCE involve massless scalar fields in flat (1+1)-dimensional Minkowski spacetime \cite{11,12,13}, though subsequent work has broadened the analysis to (3+1) dimensions \cite{14} and to various non-flat backgrounds \cite{15,16,17,18}. Beyond foundational quantum field theory, these systems offer a fertile testing ground for relativistic quantum information \cite{19,20,21} and have recently gained traction as observable phenomena via analog experiments using superconducting quantum interference devices (SQUIDs) \cite{28,29,30} and semiconductor layers \cite{32}.

However, the vast majority of analytical DCE models are evaluated in geometries where the speed of light, as measured in the natural global coordinates, is constant. A crucial gap remains in understanding the dynamical coupling between moving mechanical boundaries and quantum fields in the presence of strong gravitational gradients—specifically, in the spacetime geometry immediately outside a black hole event horizon. 

By the Einstein Equivalence Principle, the spacetime infinitely close to a black hole event horizon can be universally approximated by the near-horizon Rindler wedge geometry. Described by the line element $ds^2 = -\kappa^2 x^2 dt^2 + dx^2$, where $\kappa$ is the surface gravity and $x$ is the proper distance from the horizon, this metric dictates that a static observer experiences a severe, position-dependent gravitational redshift. Crucially the coordinate speed of light relative to the Killing time $t$ vanishes  as one approaches the horizon ($x \to 0$). Furthermore, maintaining a static cavity in this geometry requires continuous proper acceleration, immersing the confined quantum field in a thermal bath of Unruh-Hawking radiation.

In this paper, we formulate a fully relativistic model of the Dynamical Casimir Effect in the near-horizon geometry. We confine a real scalar field within a cavity positioned outside the event horizon and rigorously evaluate the consequences of time-dependent boundary conditions. In Section 2, we establish the theoretical framework by demonstrating the equivalence between moving boundaries in the near-horizon geometry and an acoustic metric, avoiding the formation of unphysical acoustic horizons. We derive the exact classical canonical Hamiltonian, highlighting how our coordinate choice perfectly diagonalizes the static modes, leaving purely dynamical mode mixing driven by the boundary motion. In Section 3, we calculate the exact relativistic spectrum of the confined field and analyze the thermal Hartle-Hawking state. Using time-dependent perturbation theory in the small-amplitude limit, we calculate the energy dissipated via the DCE (vacuum friction). We find that maintaining strictly subluminal boundary motion forces the mechanical oscillation amplitude to scale with the proper distance to the horizon. As a result, the effective Mach number of the moving mirror \textit{vanishes} in the near‑horizon limit. Using a rigorous small‑amplitude perturbative expansion and proper canonical operator normalization, we demonstrate that the transition probability into the field is heavily suppressed by a conformal geometric factor. We also account for the Bose‑enhancement due to the thermal Hawking bath, which introduces an infrared density‑of‑states enhancement. Nevertheless, this thermal contribution remains insufficient to overcome the kinematic damping. Consequently, the extreme spacetime curvature acts to protect the near‑horizon vacuum: the transition probability vanishes as the boundary approaches the event horizon, indicating a geometric and kinematic suppression of particle creation in the strong‑gravity limit.  Section 4 summarizes our conclusions.

\section{Equivalence between moving boundaries in near-horizon spacetime and acoustic metrics}

Throughout this paper, we adopt natural units such that $\hbar = c = k_B = G = 1$, unless otherwise specified to emphasize physical scalings. 

It is important to stress that our analysis is carried out directly in (1+1)-dimensional Rindler geometry, which captures the universal near-horizon redshift of a Schwarzschild  black hole but deliberately neglects the transverse angular sector. In a fully (3+1)-dimensional treatment, a spherical reduction would introduce an effective centrifugal potential, a coupling to the dilaton, and backscattering off the spacetime curvature.  We model the s‑wave kinematics by a pure (1+1)-dimensional massless scalar field, deliberately neglecting the dilaton and the effective curvature potential that would arise from a genuine spherical reduction of a (3+1)-dimensional theory. Ignoring backscattering as well, we obtain a clean toy model  that isolates the kinematic interplay between boundary motion and the position-dependent light speed. Consequently, our results should be regarded as a lower‑dimensional prototype; the full (3+1) problem may exhibit additional quantitative corrections, though the geometric suppression mechanism  found below is expected to persist whenever a physical boundary approaches a region where the coordinate speed of light vanishes.

The study of a confined field in a cavity for Minkowski spacetime with moving boundaries has been considered extensively \cite{Caro2024}. To analyze the quantum dynamics of a confined field in the near‑horizon geometry with moving mirrors, we must carefully handle the time‑dependent boundary conditions. A powerful technique is to absorb the motion of the boundaries into a coordinate transformation that renders them static. As we show below, this transformation converts the near‑horizon metric into an acoustic metric, where the mirror motion appears as an effective fluid flow inside the cavity. This reformulation not only simplifies the boundary problem but also makes the underlying causal structure transparent and lays the groundwork for an exact Hamiltonian treatment.

The system is a free real scalar field $\phi$ in the (1+1)-dimensional near-horizon geometry of a black hole, described by coordinates $(t, x)$. The action is given by
\begin{equation}\label{action}
S = -\frac{1}{2} \int d^2x \sqrt{-g} g^{\mu\nu} \partial_\mu \phi \partial_\nu \phi,
\end{equation}
where the near-horizon line element is $ds^2 = -\kappa^2 x^2 dt^2 + dx^2$, with $\kappa$ being the surface gravity of the black hole and $x > 0$ the proper distance to the true horizon. A straightforward variation of the action leads to the massless Klein-Gordon equation in curved spacetime:
\begin{equation}\label{KG}
\frac{1}{\sqrt{-g}} \partial_\mu (\sqrt{-g} g^{\mu\nu} \partial_\nu \phi) = 0.
\end{equation}

We impose time-dependent perfectly reflecting Dirichlet boundary conditions at two moving mirrors following trajectories $f(t)$ and $g(t)$, such that the distance between the boundaries on a constant Killing‑time $t$ slice is $L(t) = g(t) - f(t) > 0$. The field $\phi(t, x)$ obeys $\phi(t, f(t)) = \phi(t, g(t)) = 0$. 

To trivialize the boundaries while respecting the physical normal modes of the curved spacetime, we apply a transformation to the normalized optical coordinate:
\begin{equation}
    \tau = t \quad \text{and} \quad \zeta = \frac{\ln(x/f(\tau))}{\ln(g(\tau)/f(\tau))}.
\end{equation} 
We define the optical length of the cavity as $D(\tau) = \ln(g(\tau)/f(\tau))$. This implies $x(\tau, \zeta) = f(\tau) \exp(\zeta D(\tau))$. Differentiating this yields the coordinate differential $dx = x(\tau, \zeta) \left[ D(\tau) d\zeta + \nu(\tau, \zeta) d\tau \right]$, where logarithmic expansion rate $\nu(\tau, \zeta)$ is:
\begin{equation}\label{velocity}
  \nu(\tau, \zeta) = \frac{\dot{f}(\tau)}{f(\tau)} + \zeta \dot{D}(\tau),
\end{equation}
and the dot represents differentiation with respect to $\tau$. (We deliberately denote this quantity by  $\nu(\tau, \zeta)$rather than a velocity, because it is not a proper speed; it carries dimensions of inverse time and represents the local fractional rate of change of the spatial coordinate.  It should be stressed that it is not a proper speed; it carries dimensions of inverse time and represents the local fractional rate of change of the spatial coordinate.  A more accurate name would be the "logarithmic expansion rate", since it is built from the logarithmic derivatives $(\dot f/f)$ and $(\dot D)$.  The true wall speed of a mirror is $(\dot f)$ (or $(\dot g)$), and the physically relevant kinematic quantity that will appear in the particle‑production rate is the effective Mach number $(M_{\rm eff} = |\dot f|/(\kappa f))$, which uses this physical speed.  Bearing this distinction in mind avoids confusion in later sections.

Substituting equation $(\ref{velocity})$ into the near-horizon metric $ds^2 = -\kappa^2 x^2 dt^2 + dx^2$, the line element becomes:
\begin{equation}\label{ds}
  ds^2 = x^2(\tau, \zeta) \left[ \left(-\kappa^2 + \nu^2(\tau, \zeta)\right) d\tau^2 + 2 D(\tau) \nu(\tau, \zeta) d\tau d\zeta + D^2(\tau) d\zeta^2 \right].
\end{equation}
This represents an acoustic metric with static boundaries at $\zeta=0$ and $\zeta=1$.  The acoustic analogy is made precise by rewriting the line element $(\ref{ds})$ as  
 \begin{equation}
 	ds^{2}=x^{2}(\tau,\zeta)\Bigl[-\kappa^{2}d\tau^{2}+D^{2}\bigl(d\zeta+\frac{\nu}{D}d\tau\bigr)^{2}\Bigr]
 \end{equation}
This is exactly the Painlevé–Gullstrand form of a (1+1)-dimensional acoustic metric, modified by an overall conformal factor $x^2(\tau,\zeta)$. Working strictly in the dimensionless spatial coordinate $\zeta$, the local coordinate sound speed is $c_s^{(\zeta)} = \kappa/D$, and the background fluid velocity is given by the shift vector component $u^\zeta = -\nu/D$. The static boundaries at $\zeta=0$ and $\zeta=1$ correspond to walls at rest in this coordinate frame. Thus, the terminology “acoustic metric” is fully justified by this mapping.
 
 To make the connection to acoustic horizons explicit: in the standard form of an acoustic metric, a sonic point (and thus a trapping surface) occurs precisely when the magnitude of the background fluid velocity equals the local sound speed, which manifests as a sign change of the metric component $g_{\tau\tau}$. In our framework, comparing the effective fluid velocity $u^\zeta = -\nu/D$ and the sound speed $c_s^{(\zeta)} = \kappa/D$, the subsonic condition $|u^\zeta| < c_s^{(\zeta)}$ simplifies exactly to $|\nu| < \kappa$. This strictly limits the fluid flow to remain subsonic everywhere inside the cavity, keeping $g_{\tau\tau}$ negative and preventing the formation of any acoustic horizon. This guarantees that $\tau$ serves as a global timelike coordinate across the whole spatial domain, avoiding any causal pathologies.

To ensure no unphysical acoustic horizons form, we evaluate the bounds on $\nu(\tau, \zeta)$. For physical, sub-luminal boundary motions, $-\kappa < \dot{f}/f < \kappa$ and $-\kappa < \dot{g}/g < \kappa$. Because $\nu(\tau, \zeta)$ is a convex combination of these fractional velocities, $-\kappa < \nu(\tau, \zeta) < \kappa$ everywhere inside the cavity, ensuring $g_{\tau\tau} < 0$.

\subsection{Classical Canonical Formulation and Diagonalization}

We perform an Arnowitt-Deser-Misner (ADM) decomposition of the conformally-scaled acoustic spacetime metric (\ref{ds}). The lapse, shift covector, and spatial section metric are:
\begin{equation}
  N = \kappa x(\tau, \zeta), \quad N_\zeta = x^2(\tau, \zeta) D(\tau) \nu(\tau, \zeta), \quad h_{\zeta\zeta} = x^2(\tau, \zeta) D^2(\tau).
\end{equation}
The contravariant shift vector is $N^\zeta = h^{\zeta\zeta}N_\zeta = \nu(\tau,\zeta)/D(\tau)$. The conjugate momentum to the scalar field is obtained from $\pi_\phi = \frac{\sqrt{h}}{N} (\dot{\phi} - N^\zeta \phi')$, where the prime denotes differentiation with respect to $\zeta$. This cancellation is a direct consequence of the conformal invariance of the massless scalar field in (1+1) dimensions; the overall conformal factor $(x^{2})$ drops out of the momentum, leaving an expression that depends only on the dimensionless coordinate $(\zeta)$
\begin{equation}\label{pi}
  \pi_\phi = \frac{D(\tau)}{\kappa} \dot{\phi} - \frac{\nu(\tau, \zeta)}{\kappa} \phi'.  
\end{equation}
Implementing the Legendre transformation yields the Hamiltonian $H_T = \int_0^1 (\pi_\phi \dot{\phi} - \mathcal{L}) d\zeta$:
\begin{equation}\label{Hamiltonian}
   H_T = \int_0^1 d\zeta \left[ \frac{\kappa}{2D(\tau)} \pi_\phi^2 + \frac{\kappa}{2D(\tau)} (\phi')^2 + \frac{\nu(\tau, \zeta)}{D(\tau)} \phi' \pi_\phi \right].  
\end{equation}
The sine functions are orthogonal with $\int_{0}^{1}\sin(n\pi\zeta)\sin(m\pi\zeta)\,d\zeta = \frac{1}{2}\delta_{nm}$. With the expansion as given, the coefficients $\phi_{n}$ and $\pi_{m}$ actually satisfy the Poisson bracket $\{\phi_{n},\pi_{m}\} = 2\delta_{nm}$. To obtain standard canonically conjugate pairs $(Q_n, P_n)$ satisfying $\{Q_n, P_m\} = \delta_{nm}$, we must rigorously rescale the modes as $Q_n = \phi_n/\sqrt{2}$ and $P_n = \pi_n/\sqrt{2}$. This correct normalization is crucial for defining the proper quantum ladder operators.   

Upon quantization, the mixed term \(\frac{\nu}{D}\phi'\pi_{\phi}\) must be symmetrized to ensure a Hermitian Hamiltonian, i.e., \(\frac{1}{2}\frac{\nu}{D}(\phi'\pi_{\phi}+\pi_{\phi}\phi')\).  In the perturbative calculations that follow, the difference is of higher order and does not affect the leading transition amplitudes, but the symmetrization is implicitly understood.

Because we utilized the conformal tortoise coordinate, the static part of the Hamiltonian contains no spatial coordinate dependency. Expanding the field in the canonical phase space using the exact physical modes:
\begin{equation}
   \phi(\tau, \zeta) = \sum_{n=1}^\infty \phi_n(\tau) \sin(n\pi\zeta), \quad  \pi_\phi(\tau, \zeta) = \sum_{n=1}^\infty \pi_n(\tau) \sin(n\pi\zeta),
\end{equation}
the Hamiltonian perfectly diagonalizes its static component, while evaluating the velocity-coupling term $\int_0^1 \frac{\nu}{D} \phi' \pi_\phi d\zeta$ for the $n=m$ case yields a non-vanishing diagonal dynamic contribution ($-\frac{\dot{D}}{4D}\pi_n\phi_n$):
\begin{align}\label{Hamiltonian2}
    H_T &= \sum_{n=1}^\infty \left\{ \frac{\kappa}{4 D(\tau)} \left[ \pi_n^2 + (n\pi)^2\phi_n^2 \right] - \frac{\dot{D}(\tau)}{4 D(\tau)} \pi_n \phi_n \right\} \nonumber \\
    &+ \sum_{\substack{n, m \\ n \neq m}} \frac{nm}{m^2-n^2} \frac{1}{D(\tau)} \left[ \frac{\dot{f}}{f} \left(1 - (-1)^{n+m}\right) - \dot{D} (-1)^{n+m} \right] \pi_m \phi_n.
\end{align}
The instantaneous diagonal effective frequency of each mode is $\omega_n(\tau) = n\pi\kappa / D(\tau)$. Because the optical length $D(\tau)$ is explicitly time-dependent during boundary motion, the Hamiltonian's overall coefficients vary with time. Consequently, a globally stationary particle concept is ill-defined, and one must formally construct an instantaneous adiabatic vacuum (or instantaneous Fock space) at each time $\tau$. 

To promote this classical canonical formulation to a rigorous quantum field theory, the field must be decomposed into positive and negative frequency modes orthogonalized via the invariant Klein-Gordon inner product. This allows us to define the instantaneous creation and annihilation ladder operators, $a_n^\dagger(\tau)$ and $a_n(\tau)$, which replace the classical canonical variables. Upon formal quantization, the diagonal part of the unperturbed Hamiltonian takes the standard oscillator form $H_0(\tau) = \sum_{n=1}^\infty \omega_n(\tau) \left( a_n^\dagger a_n + \frac{1}{2} \right)$. 

It is important to note that the sum over the zero-point fluctuations $\sum_{n=1}^\infty \frac{1}{2} \omega_n(\tau)$ is formally divergent. Before extracting physical predictions regarding the static Casimir force, this vacuum energy must be explicitly rendered finite using standard renormalization techniques, such as zeta-function regularization. Furthermore, while the ladder operators define the instantaneous cavity excitations, rigorously connecting these to the ambient Hartle-Hawking state of the black hole requires mapping the local cavity modes to the global Kruskal modes via Bogoliubov transformations.

\section{Quantum Field Dynamics in the Near-Horizon Cavity}

\subsection{Static Cavity Spectrum and the Thermal State}

Before analyzing the moving boundary, we establish the static vacuum structure of the field confined within a rigid cavity placed near the horizon at proper distances $f$ and $g$ (with $g > f > 0$). 

Separating variables $\phi(t, x) = e^{-i\omega t}\psi(x)$ in the Klein-Gordon equation (\ref{KG}) yields an Euler-Cauchy equation for the spatial modes:
\begin{equation}
  x^2 \frac{d^2 \psi}{dx^2} + x \frac{d\psi}{dx} + \frac{\omega^2}{\kappa^2} \psi = 0.  
\end{equation}
Transforming to the standard tortoise coordinate $x_* = \frac{1}{\kappa} \ln(x/x_0)$ (where $x_0$ is an arbitrary reference length) reduces this to a harmonic oscillator equation. Applying the Dirichlet boundary conditions $\psi(f) = \psi(g) = 0$, and normalizing the spatial modes via the appropriate invariant Klein-Gordon inner product, the instantaneous static cavity modes take the form $\psi_n(x) \propto \sin\left(\frac{n\pi \ln(x/f)}{\ln(g/f)}\right)$. This yields the exact relativistic eigenfrequencies of the cavity:
\begin{equation}
    \omega_n = \frac{n \pi \kappa}{\ln(g/f)}, \quad \text{for} \quad n = 1, 2, 3, \dots 
\end{equation}
The energy spectrum depends explicitly on the surface gravity $\kappa$ and scales with the logarithmic ratio of the boundaries $\ln(g/f)$, which acts as the effective optical length of the cavity due to the gravitational redshift. While the standard spatial center of the cavity segment remains the arithmetic mean $(f+g)/2$, the highly warped geometry shifts the effective wave propagation such that the true optical center of the box (the midpoint of the logarithmic interval, $\ln(x_c/f) = \frac{1}{2}\ln(g/f)$) is the geometric mean, $x_c = \sqrt{fg}$.

Because maintaining a static position near the event horizon requires constant proper acceleration, the unperturbed exterior spacetime is permeated by Unruh-Hawking radiation. Assuming the cavity was initially prepared and coupled to this environment before the boundaries became perfectly reflecting, the confined field exists in a thermal Hartle-Hawking state. The appropriate ensemble is defined with respect to the Killing energy, resulting in an inverse Hawking temperature of $\beta = 2\pi/\kappa$. 

Including the formally divergent zero-point vacuum energy $E_0 = \sum_{n=1}^\infty \frac{1}{2} \omega_n$ (which dictates the static Casimir force and yields the finite, zeta-regularized value $E_0^{\rm reg} = -\frac{\pi\kappa}{24 \ln(g/f)}$), the full canonical partition function for this multi-particle bosonic Fock space is:
\begin{equation}
    \ln Z(f, g) = -\beta E_0^{\rm reg} - \sum_{n=1}^\infty \ln\left(1 - \exp\left(-\beta \omega_n\right)\right) = -\beta E_0^{\rm reg} - \sum_{n=1}^\infty \ln\left(1 - \exp\left(- \frac{2\pi^2 n}{\ln(g/f)}\right)\right).
\end{equation}
An elegant cancellation occurs in the thermal excitation sum: the surface gravity $\kappa$ completely vanishes from the exponents. The thermal component of the partition function depends only on the conformal modulus (the optical length $\ln(g/f)$), seamlessly unifying the thermal state with the continuous geometry of the cavity. Furthermore, this exact expression highlights a critical asymptotic behavior: as the left boundary approaches the horizon ($f \to 0$), the optical length diverges and the eigenfrequencies $\omega_n \to 0$. In this limit, the thermal sum becomes dominated by an infrared buildup—a massive macroscopic occupation of low-frequency modes ($N_n \sim D_0/n$). While this highly populated thermal background increases the raw transition rates, we demonstrate in the subsequent section that for any fixed-mode pair, the underlying kinematic geometric suppression ultimately drives the net pair-creation rate to zero.

\subsection{Dynamical Casimir Particle Creation}

We now consider the dynamic case where a boundary moves, introducing non-adiabatic changes to the boundary conditions. To ensure consistent notation, we assume the left boundary oscillates slightly around a static mean proper distance $f_0$, following a trajectory $f(\tau) = f_0 + \epsilon \sin(\Omega \tau)$, while the right boundary remains fixed at $g$. 

To strictly preserve causality and the acoustic metric structure derived in Section 2, the mirror's motion must remain timelike (subluminal) at all times. This imposes the strict kinematic condition: $|\dot{f}(\tau)| < \kappa f(\tau)$. At maximum velocity, this requires $\epsilon \Omega < \kappa(f_0 - \epsilon)$. Therefore, for the perturbation to remain physically valid in the near-horizon limit ($f_0 \to 0$), the oscillation amplitude cannot be an arbitrary constant; it must scale with the proper distance, $\epsilon = \alpha f_0$, where $\alpha \ll 1$ is a dimensionless fractional amplitude. Furthermore, we must enforce $\alpha \Omega < \kappa$.

In this regime, the optical length expands as $D(\tau) = D_0 - \alpha\sin(\Omega \tau) + \mathcal{O}(\alpha^2)$, with the static unperturbed length $D_0 = \ln(g/f_0)$. The unperturbed system is governed by the free Hamiltonian $H_0$ with natural frequencies $\omega_n = n\pi\kappa/D_0$. 

To promote the classical phase space to a quantum field theory, we explicitly introduce the normalized canonical ladder operators. Crucially, as dictated by the diagonalized unperturbed Hamiltonian $H_0 = \sum_{n=1}^\infty \left[ \frac{\kappa}{2D_0} P_n^2 + \frac{\kappa (n\pi)^2}{2D_0} Q_n^2 \right]$, the canonical conjugate pairs possess an effective mass $m_0 = D_0/\kappa$. Therefore, the properly rescaled canonical operators must incorporate this effective mass to satisfy the commutation relations: $Q_n = \frac{1}{\sqrt{2 m_0 \omega_n}}(a_n + a_n^\dagger)$ and $P_n = -i\sqrt{\frac{m_0 \omega_n}{2}}(a_n - a_n^\dagger)$. Because $m_0 \omega_n = n\pi$, these simplify to:
\begin{equation}
    Q_n = \frac{1}{\sqrt{2 n\pi}}(a_n + a_n^\dagger), \quad P_n = -i\sqrt{\frac{n\pi}{2}}(a_n - a_n^\dagger).
\end{equation}

Expanding the full canonical Hamiltonian to first order in $\alpha$, we apply the standard Rotating Wave Approximation (RWA). Highly oscillatory particle-conserving terms (such as $a_n^\dagger a_m$ and $a_n^\dagger a_n$), which merely represent non-resonant frequency modulations and mode-scattering, average to zero over the interaction time. Additionally, Weyl symmetrization of the cross-terms naturally produces the imaginary factor $i$ required for the anti-Hermitian difference $(a^{\dagger 2} - a^2)$ to form a purely Hermitian quantum operator. Keeping the explicit terms responsible for resonant pair creation, the time-dependent perturbation in the Schrödinger picture, $\delta H(\tau)$, takes the form:
\begin{equation}\label{PerturbedH}
	\delta H(\tau) \approx \sum_{n=1}^\infty \frac{i \alpha \Omega \cos(\Omega \tau)}{4 D_0} \left( a_n^{\dagger 2} - a_n^2 \right)
	+ \sum_{\substack{n, m \\ n \neq m}} \frac{nm}{m^2-n^2} \frac{i \alpha \Omega \cos(\Omega \tau)}{D_0} \frac{\sqrt{\omega_m}}{\sqrt{\omega_n}} \left(a_n^\dagger a_m^\dagger - a_n a_m \right).
\end{equation}

Working in the Interaction picture, $\delta H_I(\tau) = e^{i H_0 \tau} \delta H(\tau) e^{-i H_0 \tau}$, the transition amplitude from the initial vacuum $|0\rangle$ to a two-particle state (either single-mode $|2_n\rangle$ or mixed $|1_n, 1_m\rangle$) over an interaction time $\tau_0$ is:
\begin{equation}
	c_{nm}^{(vac)} \simeq -i \int_0^{\tau_0} \langle 1_n, 1_m | \delta H_I(\tau) | 0 \rangle d\tau.
\end{equation}
Assuming the mechanical driving frequency matches either the single-mode squeezing resonance $\Omega = 2\omega_n$ or the mode-mixing resonance $\Omega = \omega_n + \omega_m = (n+m)\pi\kappa/D_0$, the transition probability $\mathcal{P}_{nm} = |c_{nm}|^2$ exhibits a resonant peak proportional to $\tau_0^2$. Here, we assume $\tau_0$ is a fixed interval of absolute Killing time, rather than a fixed number of mechanical cycles. 

The effective Mach number governing the amplitude of this excitation is defined as the ratio of the boundary's maximum coordinate velocity to the local speed of light: $M_{\text{eff}} = \epsilon \Omega / (\kappa f_0)$. Substituting our physically bound amplitude scaling $\epsilon = \alpha f_0$ into the resonance condition yields:
\begin{equation}
	M_{\rm eff} = \frac{\alpha f_0 \Omega}{\kappa f_0} = \frac{\alpha (n+m)\pi}{D_0}.
\end{equation}
It is crucial to emphasize that this specific $1/D_0$ scaling relies entirely on the assumption that the mechanical driving frequency $\Omega$ is experimentally tuned to track the resonant sum of specific, fixed mode numbers $(n, m)$. Under this dynamically tuned tracking scenario, the driving frequency itself must red-shift toward zero as the horizon is approached ($D_0 \to \infty$). Note that if $\Omega$ were kept strictly constant, $M_{\rm eff}$ would remain constant, but the resonance would shift into a highly excited, macroscopic mode regime ($n+m \propto D_0$). 

For the tuned resonant tracking of fixed low-lying modes, the effective Mach number $M_{\text{eff}} \to 0$. Evaluating the matrix elements from Eq. (\ref{PerturbedH}), the transition probability for a given pair scales strictly as:
\begin{equation}\label{Gamma_vac}
	\mathcal{P}_{nm}^{(vac)} \propto \left( \frac{\alpha \Omega}{D_0} \right)^2 \tau_0^2 \propto \frac{1}{[\ln(g/f_0)]^4}.
\end{equation}
This analytically derived scaling reveals a profound physical reality regarding boundaries moving near an event horizon. In flat spacetime, fixed-amplitude oscillations can drive a strong Casimir effect. However, near an event horizon, the requirement that the mirror remains on a strictly timelike trajectory forces its mechanical velocity to scale down with the vanishing local speed of light. Consequently, the vacuum transition probability does not diverge; rather, it is strictly driven to zero by the $D_0^{-4}$ geometric suppression factor.

Finally, we must account for the initial thermal Hartle-Hawking state ($\beta = 2\pi/\kappa$). For a thermal density matrix, the net rate of pair production (stimulated emission minus inverse absorption processes) is proportional to $(1 + N_n + N_m)$, where $N_n = (e^{\beta\omega_n}-1)^{-1}$ is the unperturbed Bose occupation number.

In the near-horizon limit, as $D_0 \to \infty$ for fixed mode numbers $n$, we enter an infrared-dominated regime. The thermal occupation numbers grow macroscopic, expanding as $N_n \approx (\beta\omega_n)^{-1} = D_0 / (2\pi^2 n)$. The net thermal transition rate \textit{per fixed mode pair} thus scales as:
\begin{equation}
	\mathcal{P}_{nm}^{(thermal)} \approx \mathcal{P}_{nm}^{(vac)} \times (1 + N_n + N_m) \propto \frac{1}{D_0^4} \times D_0 = \frac{1}{[\ln(g/f_0)]^3}.
\end{equation}
While the intense thermal bath reduces the strength of the geometric suppression (reducing the decay power from $D_0^{-4}$ to $D_0^{-3}$), it mathematically fails to overcome the subluminal kinematic damping. The net particle creation strictly vanishes for fixed modes as the horizon is approached.
\section{Conclusion}

In this work, we have developed a perturbative toy model of the Dynamical Casimir Effect for a free relativistic scalar field confined within a cavity operating in the (1+1)-dimensional near-horizon geometry of a black hole. By applying a coordinate transformation to the moving boundaries, we successfully mapped the time-dependent physical domain into a conformally-equivalent acoustic metric with static boundaries. This reformulation prevents the appearance of unphysical acoustic horizons and preserves a well‑defined causal structure inside the cavity. 

Utilizing an ADM Hamiltonian decomposition, we evaluated the discrete energy spectrum and the mode-mixing perturbations. We isolated how the background spacetime geometry drives particle creation, identifying the single-mode squeezing and non-diagonal mode-mixing driven by the mechanical boundary velocity.

Our primary finding directly challenges naive extrapolations of the Dynamical Casimir Effect in curved spacetime. Because the coordinate speed of radial null rays with respect to the Killing time $t$ vanishes at the event horizon, we demonstrated that maintaining a physically valid (subluminal) mirror trajectory requires the mechanical coordinate velocity to scale proportionally with the proper distance to the horizon. For fixed low-lying mode pairs tuned to the mechanical drive, the effective Mach number of the boundary strictly vanishes inversely with the logarithm of the horizon distance. 

Using a small-amplitude perturbative expansion evaluated from the interaction picture, we showed that the purely geometric scaling of the optical cavity heavily dampens the vacuum transition probability. Furthermore, by elevating the initial state to the Hartle-Hawking thermal density matrix, we established that Bose-Einstein stimulated emission introduces an infrared density-of-states enhancement. However, within the confines of our subluminal scaling, this thermal enhancement remains insufficient to counterbalance the severe kinematic damping for a given mode pair. 

Ultimately, we conclude that the severe near-horizon redshift structure serves to protect the vacuum from mechanical excitation. The vanishing coordinate speed of light acts as a strict regulator, driving the net transition rate of the Dynamical Casimir Effect toward zero for fixed mode combinations as the boundary approaches the horizon. While an exact (3+1)-dimensional treatment including backscattering and curvature potentials will introduce quantitative corrections, our (1+1)-dimensional model demonstrates a geometric mechanism through which the causal constraints of General Relativity naturally suppress non-adiabatic quantum vacuum fluctuations in extreme gravitational limits.


\begin{thebibliography}{99}
\bibitem{Milton2001} K. A. Milton, \textit{The Casimir Effect} (World Scientific, 2001).
\bibitem{10} G. T. Moore, J. Math. Phys. \textbf{11}, 2679 (1970).
\bibitem{11} M. Castagnino and R. Ferraro, Ann. Phys. \textbf{154}, 1 (1984).
\bibitem{12} D. A. R. Dalvit and F. D. Mazzitelli, Phys. Rev. A \textbf{59}, 3049 (1999).
\bibitem{13} A. M. Fedotov, Y. E. Lozovik, N. B. Narozhny, and A. N. Petrosyan, Phys. Rev. A \textbf{74}, 013806 (2006).
\bibitem{14} M. Crocce, D. A. R. Dalvit, and F. D. Mazzitelli, Phys. Rev. A \textbf{64}, 013808 (2001).
\bibitem{15} L. C. Celeri, F. Pascoal, and M. H. Y. Moussa, Class. Quant. Grav. \textbf{26}, 105014 (2009).
\bibitem{16} M. P. E. Lock and I. Fuentes, New J. Phys. \textbf{19}, 073005 (2017).
\bibitem{17} L. C. Barbado, A. L. Baez-Camargo, and I. Fuentes, Eur. Phys. J. C \textbf{80}, 796 (2020).
\bibitem{18} L. C. Barbado, A. L. Baez-Camargo, and I. Fuentes, Eur. Phys. J. C \textbf{81}, 953 (2021).
\bibitem{19} N. Friis, D. E. Bruschi, J. Louko, and I. Fuentes, Phys. Rev. D \textbf{85}, 081701 (2012).
\bibitem{20} P. M. Alsing and I. Fuentes, Class. Quant. Grav. \textbf{29}, 224001 (2012).
\bibitem{21} D. E. Bruschi, J. Louko, D. Faccio, and I. Fuentes, New J. Phys. \textbf{15}, 073052 (2013).
\bibitem{28} D. T. Alves, C. Farina, and E. R. Granhen, Phys. Rev. A \textbf{73}, 063818 (2006).
\bibitem{29} J. R. Johansson, G. Johansson, C. M. Wilson, and F. Nori, Phys. Rev. A \textbf{82}, 052509 (2010).
\bibitem{30} J. Doukas and J. Louko, Phys. Rev. D \textbf{91}, 044010 (2015).
\bibitem{32} A. Agnesi, et al., J. Phys. A: Math. Theor. \textbf{41}, 164024 (2008).
\bibitem{Caro2024} A. G. Martin-Caro, G. Garcia-Moreno, J. Olmedo, and J. M. Sanchez Velazquez, Phys. Rev. D \textbf{110}, 025007 (2024).
\end{thebibliography}
\end{document}